\documentclass[12pt,preprint]{emulateapj}

\newcommand{\gb}{$g'$}
\newcommand{\rb}{$r'$}
\newcommand{\ib}{$i'$}
\newcommand{\zb}{$z'$}
\newcommand{\alp}{$\alpha$ }
\newcommand{\ang}{$\mbox{\AA}$}
\newcommand{\sol}{$_{\odot}$ }
\newcommand{\super}[1]{$^{#1}$}
\newcommand{\sub}[1]{$_{#1}$}
\newcommand{\lya}{Ly$\alpha$}
\def\farcs{\hbox{$.\!\!^{\prime\prime}$}}

\shorttitle{\lya~Galaxies at z $\sim$ 4.5}
\shortauthors{Finkelstein et al.}

\begin{document}
\title{The Ages and Masses of Lyman Alpha Galaxies at z $\sim$ 4.5\altaffilmark{1}}

\author{Steven   L.    Finkelstein\altaffilmark{2},   James   E.    Rhoads,   Sangeeta
Malhotra\altaffilmark{3},  Norbert Pirzkal\altaffilmark{4}  \& Junxian
Wang\altaffilmark{5}}   
\affil{$^{1}$  Observations reported  here  were obtained  at the  MMT
Observatory, a  joint facility  of the University  of Arizona  and the
Smithsonian Institution}
\affil{$^{2}$  Department  of  Physics,  Arizona  State
University,  Tempe, AZ  85287} 
\affil{$^{3}$  School of Earth and Space Exploration,  Arizona  State
University,  Tempe, AZ  85287} 
\affil{$^{4}$  Space  Telescope Science
Institute, 3700  San Martin Drive, Baltimore,  MD 21218} 
\affil{$^{5}$ University of  Science and Technology  of China, Hefei,  Anhui 230026,
China}

\begin{abstract}
  
  We examine  the stellar populations of  a sample of 98  z $\sim$ 4.5
  Lyman alpha  emitting galaxies using their  broadband colors derived
  from deep  photometry at the  MMT.  These galaxies were  selected by
  narrowband  excess   from  the   Large  Area  Lyman   Alpha  survey.
  Twenty-two galaxies are  detected in two or more  of our MMT filters
  (g', r',  i' and z').  By  comparing broad and  narrowband colors of
  these galaxies  to synthetic colors from  stellar population models,
  we determine  their ages and  stellar masses.  The highest  equivalent width
  objects have an  average age of 4 Myr,  consistent with ongoing star
  formation.   The lowest  EW objects  show an  age of  40 -  200 Myr,
  consistent  with the expectation  that larger  numbers of  stars are
  causing low EWs.  We  found masses ranging from 2$\times$10\super{7}
  M\sol for the youngest objects in the sample to 2$\times$10\super{9}
  M\sol  for the  oldest.   It  is possible  that  dust effects  could
  produce  large  equivalent  widths  even  in  older  populations  by
  allowing the Lyman alpha photons to escape, even while the continuum
  is extinguished, and we present models for this scenario also.

\end{abstract}


\section{Introduction}

There  are  two  popular  techniques  for locating  galaxies  at  high
redshift:  the Lyman  Break technique  \citep{steidel}  and narrowband
selection.   The second  entails observing  a galaxy  in  a narrowband
filter containing  a redshifted emission line.  This  method is useful
to  select  high redshift  galaxies  with  strong  Lyman alpha  (\lya)
emission  lines.  While the  line emission  properties of  Lyman Break
galaxies  (LBGs)  are  known,  little  is known  about  the  continuum
properties of  \lya~galaxies because they  are so faint.  Studying the
continuum properties of these  \lya~galaxies at high-redshift can tell
us about their age, stellar mass, dust content and star formation history.

We have  used the narrowband  selection technique to locate  and study
Lyman alpha emitting galaxies (LAEs) at z $\sim$ 4.5.  Similar studies
have been done both in this field and in other fields (e.g., Rhoads et
al 2000, 2004; Rhoads \& Malhotra 2001; Malhotra \& Rhoads 2002; Cowie
\& Hu  1998; Hu et al 1998,  2002, 2004; Kudritzki et  al 2000; Fynbo,
Moller, \& Thomsen 2001; Pentericci  et al 2000; Stiavelli et al 2001;
Ouchi et al 2001, 2003, 2004; Fujita et al 2003; Shimasaku et al 2003,
2006; Kodaira  et al  2003; Ajiki  et al 2004;  Taniguchi et  al 2005;
Venemans et al 2002, 2004).  In many of these studies, the strength of
the Lyman  alpha (\lya) line  has been found  to be much  greater than
that of a normal stellar  population (Kudritzki et al. 2000; Dawson et
al. 2004).  Malhotra and  Rhoads (2002; hereafter MR02) found numerous
LAEs with rest frame equivalent widths (EWs) $>$ 200~\ang.  Assuming a
constant star  formation rate, the  EW of a normal  stellar population
will  asymptote  toward 80 ~\ang~by  10\super{8}  years,  down from  a
maximum  value of  $\sim$ 240 ~\ang.  These  galaxies are  of interest
because it  is believed  that this strong  \lya~emission is a  sign of
ongoing star formation activity in these galaxies \citep{pp67}, and it
could mean that  these are some of the youngest  galaxies in the early
universe.

There are  a few possible scenarios  that could be creating  this large
EW.  Strong \lya~emission could  be produced via star formation if the
stellar photospheres are hotter than normal, which could happen in low
metallicity  galaxies.  It could  also mean  that these  galaxies have
their stellar  mass distributed via a top-heavy  initial mass function
(IMF).  Both  of these scenarios  are possible in  primitive galaxies,
which are thought to contain young stars and little dust.

Active galactic  nuclei also produce  a large \lya~EW.  In  an optical
spectroscopic  survey of  z $\sim$  4.5 \lya~galaxies,  Dawson  et al.
(2004) found large \lya~equivalent  widths, but narrow physical widths
($\Delta$v  $<$ 500 km  s\super{-1}).  They  also placed  tight upper
limits  on the  flux  of accompanying  high-ionization state  emission
lines (e.g.,  NV $\lambda$1240, SiIV  $\lambda$1398, CIV $\lambda$1549
and HeII $\lambda$1640), suggesting  that the large \lya~EW is powered
by star formation rather than AGN.  Additionally, Type I (broad-lined)
AGN are ruled out because the width of the \lya~line in such an object
is greater  than the width of  the narrowband filters  which were used
(see  Section 2.3).  While  Type II  AGN lines  are narrower  than our
filters,  they remain  broader  than  the typical  line  width of  our
\lya~sample as  determined by  followup spectroscopy (e.g.,  Rhoads et
al.  2003).  Malhotra et al.   (2003) and Wang et al.  (2004) searched
for a correlation between LAEs and Type II AGN using deep {\it Chandra
  X-Ray Observatory}  images.  101 \lya~emitters were known  to lie in
the  two fields observed  with {\it  Chandra}, and  none of  them were
detected  at  the  3$\sigma$  level  (L\sub{2-8 keV}  =  2.8  $\times$
10\super{42} ergs s\super{-1}).   The sources remained undetected when
they  were stacked, and  they concluded  that less  than 4.8\%  of the
\lya~emitters they studied could  be possible AGNs.  Therefore, we are
confident that there are at most a few AGN in our \lya~galaxy sample.

Another possible  scenario involves enhancement of the  \lya~line in a
clumpy medium.  In this scenario, the continuum light is attenuated by
dust whereas  the \lya~line  is not, resulting  in a  greatly enhanced
observed \lya~EW (Neufeld  1991; Hansen \& Oh 2006).   This can happen
if  the  dust  is  primarily  in cold,  neutral  clouds,  whereas  the
inter-cloud medium  is hot  and mainly ionized.   Because \lya~photons
are  resonantly scattered,  they  are preferentially  absorbed at  the
surface.  Thus it is highly unlikely that a \lya~photon will penetrate
deep into one  of these clouds.  They will  be absorbed and re-emitted
right at  the surface,  effectively scattering off  of the  clouds and
spending  the  majority  of  their  time in  the  inter-cloud  medium.
However, continuum  photons are not  resonantly scattered, and  if the
covering factor of clouds is  large, they will in general pass through
the interior of  one.  Thus the continuum photons  will suffer greater
attenuation than the \lya~photons, effectively enhancing the \lya~EW.

These \lya~galaxies are so faint compared to other high-z objects that
have  been studied,  that not  a lot  is known  about  their continuum
properties.   Our primary  goal is  to better  constrain the  ages and
stellar masses of these objects from  their continuum light in order to better
understand the driving force behind the strength of the \lya~line.  We
have obtained deep broadband images in the \gb, \rb, \ib~and \zb~bands
for this  purpose.  Using the colors  of these galaxies,  we can study
continuum  properties  of individual  \lya~emitting  galaxies at  this
redshift for the first time,  allowing us to estimate their age, mass,
star formation rate  and dust content.  The next  phase of our project
will be to look into the likelihood of dust causing the large \lya~EW,
which we will begin to do in this paper, as the colors of these galaxies
might  distinguish the  cause  of  the large  EW.   Blue colors  would
indicate  young stars with  hot photospheres,  while red  colors would
indicate  dust  quenching of  the  continuum,  enhancing the  \lya~EW.
Observations,  data  reduction  technique  and  sample  selection  are
reported in  Section 2; results are  presented in Section  3, and they
are discussed in Section 4.  Conclusions are presented in Section 5.

\section{Data Handling}

\subsection{Observations}
The Large  Area Lyman Alpha (LALA)  survey began in 1998  using the 4m
Mayall telescope  at the Kitt  Peak National Observatory  (KPNO).  The
final  area  of this  survey  was  0.72  deg\super{2} in  two  fields,
Bo$\ddot{o}$tes         and         Cetus,         centered         at
14\super{h}25\super{m}57\super{s},   +35\super{o}32$'$   (J2000)   and
02\super{h}05\super{m}20\super{s},      -04\super{o}55$'$      (J2000)
respectively  \citep{rh00}.  The  LALA survey  found large  samples of
\lya~emitting  galaxies at  z  =  4.5,  5.7  and 6.5,  with  a
spectroscopic success rate of up  to 70\%.  We observed the LALA Cetus
Field for three  full nights in 2005 November, and  again for four 1/4
nights in  2006 January, using the Megacam  instrument \citep{mega} at
the Multiple  Mirror Telescope (MMT).   Megacam is a large  mosaic CCD
camera with a 24'x24' field of view, made up of 36 CCDs with 2048x4608
pixels.   Each  pixel is  0\farcs08  on  the  sky.  We  acquired  deep
broadband images in four Sloan Digital Sky Survey (SDSS) filters: \gb,
\rb, \ib~and  \zb.  The total exposure  time in the  \gb, \rb, \ib~and
\zb~filters were 4.33, 3.50,  4.78 and 5.33 hours respectively.  Given
that the seeing during the run  was rarely below 0\farcs8, we chose to
bin  the pixels  2$\times$2 to  reduce data  volume, resulting  in a
final pixel scale of $\sim$0\farcs16 per pixel in our individual images.

\subsection{Data Reduction}
The data  were reduced in IRAF\footnote[1]{IRAF is  distributed by the
National Optical Astronomy Observatories  (NOAO), which is operated by
the  Association  of Universities  for  Research  in Astronomy,  Inc.\
(AURA)   under  cooperative  agreement   with  the   National  Science
Foundation.}      \citep{tody86,tody93},      using     the     MSCRED
\citep{valtod,val98} and MEGARED  \citep{obs} reduction packages.  The
data were first processed using  the standard CCD reduction steps done
by the  task ccdproc (overscan correction, trim,  bias subtraction and
flat fielding).  Bad  pixels were flagged using the  Megacam bad pixel
masks included with the MEGARED package.

Fringing was  present in the \ib~and  \zb~band data, and  needed to be
removed.    For  this,   we   made  object   masks   using  the   task
objmasks\footnote[2]{In order to prevent  fringing from making it onto
the object  masks, we first made  a skyflat without  object masks, and
divided this skyflat out of the  science data.  We then got the object
masks  from  these  quasi-fringe  removed  images.  The  rest  of  the
procedure used the  untouched science frames.}.  We then  took a set of
science frames  in each  filter, and combined  them to make  a skyflat
using sflatcombine (using the object masks to exclude any objects from
the  resultant image).   This skyflat  was then  median smoothed  on a
scale of 150$\times$150 pixels  (\ib) and 200$\times$200 pixels (\zb).
The smoothed images were subtracted from the unsmoothed images, and the
result  was a  fringe  frame for  each  band.  The  fringing was  then
removed from the science images  using the task rmfringe, which scales
and subtracts the fringe  frame.  To make the illumination correction,
skyflats were made, median  smoothed 5$\times$5 pixels, normalized and
divided out of the science images.

The  amplifiers in  the  reduced images  were  merged (from  72 to  36
extensions) using the MEGARED task  megamerge.  The WCS written in the
image header at the telescope  is systematically off by 0.5 degrees in
rotation, so we  used the task fixmosaic to  adjust the rotation angle
so that  the WCS solution works  better.  The WCS  was then determined
using  megawcs, and  the WCS  distortion terms  were installed  in the
header using the task zpn.

The 36 chips  were combined into a single  extension using the program
SWarp\footnote[3]{SWarp  is  a  program  that  resamples  and  co-adds
  together  FITS   images,  authored   by  Emmanuel  Bertin.    It  is
  distributed  by Terapix  at:  http://terapix.iap.fr/soft/swarp.}. We
ran SWarp on each input  image, using a corresponding weight map which
gave zero weight to known bad  pixels and cosmic ray hits.  We located
the cosmic rays  using the method of Rhoads  (2000).  We specified the
center  of the  output images  to be  the average  center of  the five
dither    positions   we    used   (02\super{h}04\super{s}56\super{m},
-05\super{o}01$'$01$''$ J2000), which corresponded to an image size of
11988$\times$10881 pixels and a  pixel scale of 0\farcs1587 per pixel.
During  this process, each  image was  resampled (using  the Lanczos-3
6-tap filter)  and interpolated onto  the new pixel grid.   While this
was done, SWarp determined and subtracted the background value in each
extension.

An  initial stack  for each  filter was  made using  SWarp  to average
together  the individual  images.  To  make the  final version  of the
stack, we  needed to  remove the satellite  trails which  littered our
images.   To do  this,  we  created a  difference  image between  each
individual input  image and the  initial stack.  Using  a thresholding
technique, we flagged the satellite trails in this image, and combined
those flags with the existing  weight-maps.  We ran SWarp a final time
with this as  the input weight map, creating the  final stack.  We did
this process for each band, giving  us four images, one final stack in
each of  the \gb, \rb, \ib~and  \zb~bands.  In an  attempt to increase
the signal of our objects, we  tried co-adding $R$ and $I$ band images
from  the NOAO  Deep  Wide-Field Survey  (NDWFS)\citep{jdey99} to  our
\rb~and  \ib~ images,  however the  additional signal  did not  make a
significant difference in the results.

To calculate  the photometric zeropoint  for our observations  we took
images of  the standard  stars SA95~190 and  SA95~193 during  our run,
where the zeropoint is defined as  the magnitude of an object with one
count in the  image (for an integration time  of 400s). These standard
stars  are  from Landolt  (1992),  and  we  used the  transforms  from
Johnson-Morgan-Cousins  to  SDSS   magnitudes  from  Fukugita  et  al.
(1996).  Averaging  the results from the  two stars for  each band, we
obtained  these   zeropoints:  \gb=33.13,  \rb=33.00,   \ib=32.58  and
\zb=31.43.  Using these  zeropoints with our data, all  of our results
are in AB magnitudes\footnote[4]{For more details on our reduction and
  analysis process, see: http://stevenf.asu.edu/Reduction.html}.

\subsection{Lyman \alp Galaxy Selection}

To extract the objects from each stack, we used the SExtractor package
\citep{ba96}.  For SExtractor to accurately estimate the errors in the
flux, it  needed to know the  gain in the input  images.  To calculate
this, we needed  the mean and standard deviation  of the background of
the images.   Because the background  was subtracted out in  SWarp, we
obtained the  average background value  in each band by  averaging the
values  obtained from the  pre-SWarped images,  and adding  this value
onto the final  stacks.  We then ran SExtractor  in the two-image mode
using a nine pixel aperture (2.$''$32) with the narrowband images from
the LALA survey, giving us  catalogs of objects which were detected in
both images.

The method  we have  used to locate  \lya~galaxies involves  taking an
image using a narrowband  filter containing the wavelength for \lya~at
a certain redshift.  We used the catalog from MR02, which was based on
LALA narrowband images and broad  band $B_{w}$, $R$, and $I$ data from
the NDWFS (Jannuzi \& Dey 1999).  The narrowband data consists of five
overlapping narrowband  filters, each with a FWHM  $\sim$ 80\ang.  The
central  wavelengths are  $\lambda\lambda$6559 (H0),  6611  (H4), 6650
(H8), 6692 (H12) and 6730 (H16), giving a total redshift coverage 4.37
$<$ z  $<$ 4.57.  To  ensure that there  was no overlap, we  only used
objects selected from the H0, H8  and H16 filters in our analysis.  In
order to compare  the two data sets, the  Megacam data were registered
and remapped onto the same scale as the narrowband data using the IRAF
tasks wcsmap and geotran.

Selection criteria for the  MR02 catalog, following Rhoads \& Malhotra
2001, were  as follows: (1)  5$\sigma$ significance detection  in the
narrowband: This is calculated  using an SExtractor aperture flux with
the  associated  flux  error;  flux/error  $\geq$  5.   (2)  4$\sigma$
significant  excess of narrowband  flux: This  was calculated  by taking
narrowband flux  - broadband flux, both calibrated  in physical units,
and    the   associated   error    (sqrt(error\super{2}\sub{broad}   +
error\super{2}\sub{narrow}))  and  demanding  that flux  difference  /
error $\geq$ 4.  (3) Factor of $\geq$ 2 ratio between broad and narrow
band  fluxes,  calibrated to  the  same units.   (4)  No  more than  2
$\sigma$ significant flux in the bluest filter observed ($B_{w}$).

\subsection{Stellar Population Models}

In order  to study  the properties  of the galaxies  in our  sample we
compare  them   to  stellar  population  models,   using  the  stellar
population modeling software by Bruzual \& Charlot (2003).  With these
models, we were able to choose a range of ages, metallicities and star
formation rates for comparison with our sample.  We chose ages ranging
from 10\super{6}  - 10\super{9} years, metallicities from  .02 Z\sol -
Z\sol  and   exponentially  decaying  star  formation   rates  with  a
characteristic    time-scale    of     $\tau$    =    10\super{3}    -
2$\times$10\super{9} years.   We included  dust via the  Calzetti dust
extinction  law  (Calzetti  et  al.  1994),  which  is  applicable  to
starburst galaxies,  in the  range: A\sub{1200} =  0 - 2.   Lastly, we
included intergalactic medium (IGM) absorption via the prescription of
Madau (1995).

The Bruzual \& Charlot code (BC03)  output the flux of a given stellar
population in units of L\sol\ang\super{-1}.   In order to go from flux
in these units to bandpass averaged fluxes, we used the method outlined
by Papovich et  al.  (2001).  In short, we took  the output from BC03,
and    converted     it    from    f\sub{\ensuremath{\lambda}}    into
f\sub{\ensuremath{\nu}}   (units  of   erg   s\super{-1}  cm\super{-2}
Hz\super{-1}).   This flux  was  then multiplied  by the  transmission
function for a given bandpass, and integrated over all frequency.  The
bandpass  averaged  flux  $\left<f_{\ensuremath{\nu}}\right>$ is  this
result normalized  to the integral  of the transmission  function.  AB
magnitudes \citep{oke} for the candidates were then computed.

In order  to calculate colors which  we could directly  compare to our
observations, it was  necessary to add in emission  line flux from the
\lya~line, which appears in the \rb~filter at z $\sim$ 4.5.  While the
BC03 software does not calculate  emission line strengths, we were able
to use one of its many other output quantities: the number of ionizing
photons.   We  calculated  the  Ly$\alpha$ emission  line
strength (in units of L\sol\ang\super{-1} to match the model output) by using:
\begin{equation}
Ly\alpha~Linestrength = \frac{hc}{\lambda_{Ly\alpha}L_{\odot}\Delta\lambda}\times\frac{2}{3}\times n_{ion}
\end{equation}
where n\sub{ion} is the  number of ionizing photons, $\Delta$$\lambda$
is the bin size of the wavelength array (1\ang), and the factor of 2/3
represents  the  fraction  of  ionizing  photons  which  will  produce
Ly$\alpha$ photons when interacting with the local interstellar medium
(ISM) under Case  B recombination\footnote[5]{This calculation follows
  the simple assumptions that no ionizing photons escape, and that all
  Ly$\alpha$  photons escape.}.   In order  to model  the  clumpy dust
scenario, we needed to ensure  that the Ly$\alpha$ line did not suffer
dust attenuation.   To do this,  the continuum flux was  multiplied by
the Calzetti  dust law before we  added in the Ly$\alpha$  flux to the
spectrum at the  correct wavelength bin.  In this  case, the continuum
suffers dust attenuation while the \lya~line does not.

\section{Results}

\subsection{Ly$\alpha$ Galaxy Candidates}
Using the  catalog described above,  we have identified 98  objects as
Ly$\alpha$         galaxy         candidates        within         the
24$^{\prime}$$\times$24$^{\prime}$  Megacam field  of  view.  We  have
classified 22  of these objects  as continuum detections on  the basis
that they have at least 2$\sigma$ detections in two of the \rb, \ib~or
\zb~bands (the other  76 objects were undetected at  this level in the
broadband data, but we stacked their fluxes for analysis).

We have in  our possession IMACS spectra of objects  in the LALA Cetus
Field (Wang  et al. 2007).  These  spectra have been  used to identify
the redshift of the Ly$\alpha$ line  in these galaxies.  Out of the 98
total  Ly$\alpha$ galaxy  candidates, 28  have  been spectroscopically
confirmed  as being  Ly$\alpha$ emitters  at z  $\sim$ 4.5.   Seven of
these  confirmations are  among  the 22  Megacam detections.   Another
seven of the  Megacam detections have IMACS spectra  but had no strong
Ly$\alpha$ line identified, however some of these may show a line with
deeper spectra.

\subsection{Equivalent  Width    Distribution}   
We calculated  the rest-frame equivalent  widths of our  22 detections
using the ratio of the line flux to the continuum flux via:
\begin{equation}
EW = \left(\frac{1 - \eta}{\frac{\eta}{\Delta\lambda_{BB}} - \frac{1}{\Delta\lambda_{NB}}}\right) \times (1 + z)^{-1}
\end{equation}
where $\eta$ is the ratio of narrowband to broadband flux (in units of
f\sub{\nu}),  and   the  $\Delta$$\lambda$s  are   the  filter  widths
(80\ang~and 1220\ang~for  the narrow and  broad filters respectively).
The  narrowband fluxes  are  measured from  the  LALA narrowband  data
discussed above,  and the broadband  fluxes are measured from  our MMT
\rb~data.

Figure 1 shows  the calculated  rest-frame  EWs of  our 22 continuum detected
galaxies.  The  range of EWs is 5 -- 800\ang\footnote[6]{However, the
object  with an  EW of  5~\ang~is  likely an  interloper.  It  has an
$\sim$ 10~$\sigma$ detection in \gb, and $>$ 30~$\sigma$ detections in
\rb, \ib  and \zb.  We will exclude  this object from the  rest of our
analysis,  giving us  21  detections.}.   While the  range  of our  EW
distribution peaks at  the low end of previous  studies (i.e. most of our objects
have  rest-frame EW  $\lesssim$  200\ang), we  recognize  that the  76
undetected objects would  all lie at the high EW end  of this plot (in
fact, they peak at EW well over 200\ang).  Because they are undetected, this
means that  their broadband fluxes were  too faint to  be detected in
our survey.  However, they  still have very bright narrowband fluxes,
indicating strong  \lya~emission.  This strong  emission, coupled with
faint broadband \rb~magnitudes, results in a high \lya~EW.

From  models of normal  stellar populations  (those with  a near-Solar
metallicity and  a Salpeter IMF), it  was found that the  maximum EW a
galaxy could have  is $\sim$ 240\ang~at t $\sim$  10\super{6} yr, with
the EW approaching a steady value of 80\ang~at t $\sim$ 10\super{8} yr
(MR02;  Charlot \&  Fall 1993).   Our data  show many  objects  with a
rest-frame EW greater than 100\ang.  While individual galaxies with EW
this high are  possible from normal stellar populations,  the ratio of
high-low EWs is much greater than  it ought to be when we account for
the 76 objects without MMT detections.  This implies that something is
causing the  EW to be  higher than normal.   One of our goals  in this
study is  to investigate whether  the cause is massive  stars creating
more  \lya~photons,  or  dust   suppressing  the  continuum  and  thus
enhancing the \lya~EW.
\vskip 0.25cm
\epsfxsize 3.3in
\epsffile{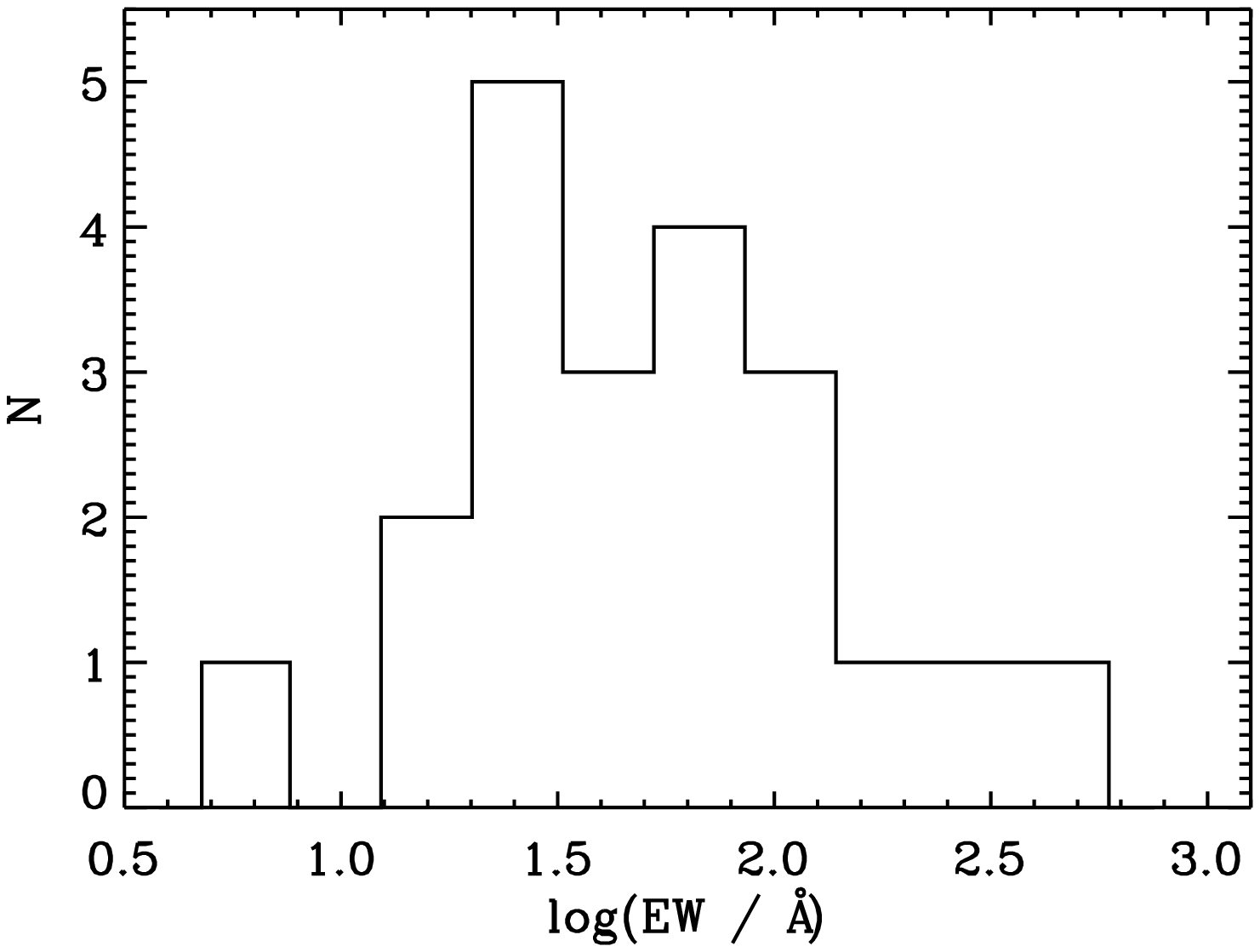}
\figcaption[f1.eps]{The  distribution of  rest-frame \lya~equivalent  widths from
  our 22  detections.  The EWs range  from 5 --  800~\ang.  The single
  object with EW $<$ 20~\ang~ is likely an interloper that has managed
  to satisfy our selection  criteria.  A normal stellar population has
  an  EW that  asymptotes toward  $\sim$  80~\ang~as its  age goes  to
  10\super{8} years,  so something is  causing the \lya~EW of  many of
  these objects to be higher..}
\vskip 18pt

We  note that  four objects  among  our MMT  detections show  observed
equivalent  width in excess  of 230\AA.   Such high  equivalent widths
might  be explained  by  a  combination of  extreme  youth and/or  low
metallicity (e.g.  Malhotra \& Rhoads 2004), or might  be produced by  radiative transfer
effects  (Neufeld 1991,  Hansen \&  Oh 2006).   However, two  of these
objects are consistent with  200\AA\ equivalent width at the 1$\sigma$
level, and  the other two at  the 2$\sigma$ level,  so such mechanisms
are not strictly required to explain these objects.  Stronger evidence
for such  effects may  be found among  the 76 sources  whose continuum
emission was too faint for our MMT data.

\subsection{Stacking Analysis}
Our goal has been  to be able to compare our objects  to the models to
obtain estimates of physical parameters  such as age and stellar mass,
along with dust content.  We divided our objects into six groups which
we then  stacked (in order to  damp down the errors)  to get composite
fluxes  which  we  could  compare  to the  models.   The  groups  are:
non-detections, which  consists of the  76 selected objects  which did
not have  a 2 $\sigma$  detection in at  least two bands (out  of \rb,
\ib~and \zb);  low EW,  which consists of  the seven  detected objects
with EW of 20--40\ang; middle EW, which consists of the seven detected
objects with EW  of 45--100\ang; high EW, which  consists of the seven
detected objects with EW $\geq$ 110\ang; detections, which consists of
the   21  objects   which   were  detected   in   the  Megacam   data;
spectroscopically  confirmed,  which consists  of  the  28 objects  (7
detected) which  were confirmed to  be \lya~emitters at z  $\sim$ 4.5.
These stacks are shown in Figure 2.

\subsection{Age and Mass Estimates}
In order to  study these galaxies, we have  created numerous synthetic
stellar population  models which we compare to  our observed galaxies.
The  most illuminating  way in  which we  can study  the  observed vs.
model galaxies are  in a color-color plot (Figures 2  and 3).  We have
plotted the \rb~- \ib~ color vs.  the \rb~- $nb$ color of our objects,
and then overplotted many model curves.  Because changing
the metallicity did  not much change the position  of the model curves
in the plane we are studying,  we have elected to only use models with
Z = .02 Z\sol.  The  differing line styles represent the two different
star  formation  rates we  used,  with  the  solid lines  representing
continuous   star  formation,  and   the  dashed   lines  representing
exponentially decaying  star formation with  a decay time of  $\tau$ =
10\super{7} years.

We  ran the  models  at  24 different  ages  ranging from  1  Myr -  2
Gyr\footnote[7]{At 1 Myr, the  two different star formation rates have
identical  colors, so  their  model  tracks overlap,  that  is why  it
appears as if  there is no curve for  an exponentially decaying galaxy
at 1 Myr.}, and in the figure we show the five ages which best surround
the data,  represented by different  colors.  We included dust  in the
models, ranging  from A\sub{1200} =  0 - 2.   The extent of  the model
tracks represents  the different dust optical depths.   As we discussed
above, this dust does not  affect the strength of the model \lya~line.
We  have connected  the zero-dust  end of  the model  tracks  with the
dotted  black line  in order  to show  where galaxies  with  no dust
should lie.
\vskip 0.25cm
\epsfxsize 3.3in
\epsffile{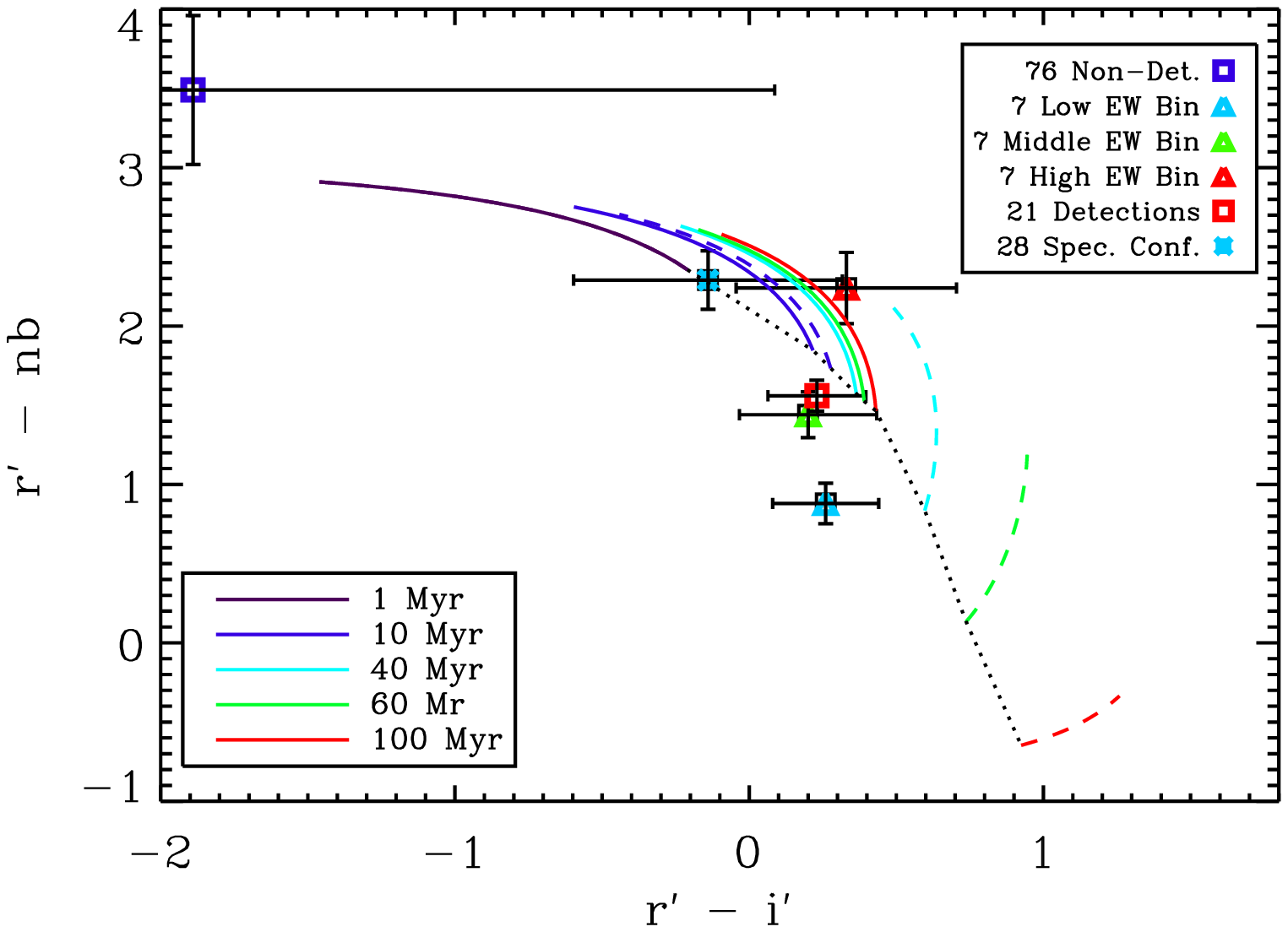}
\figcaption[f2.eps]{Color-color plot  showing the  six stacked points  along with
  stellar  population model tracks.   The vertical  axis is  the \rb~-
  narrowband  (H0)  color,  and  the  horizontal  axis  is  the  \rb~-
  \ib~color.  The model tracks shown represent the colors of models in
  this plane.   All models  are for .02  Z\sol.  The  solid tracks
  denote  continuous star  formation,  while the  dashed tracks  denote
  exponentially decaying star formation,  with a decay time of $\tau$
  = 10\super{7}  years.  The different colors  of the model curves  represent the
  ages  of the models.   The length  of the  model curves  represent the
  model colors with  differing amounts of dust.  Models  were run with
  A\sub{1200}  = 0  -- 2,  with the  A\sub{1200} point  lying  at the
  smallest  \rb~- $nb$  color.   The dotted  black  line connects  the
  bottom of the  model curves, and as such  is the ``zero-dust'' line.
  Ages were found  for each stack by minimizing  the distance from the
  stack to the nearest A\sub{1200} = 0 model point (i.e. nearest point
  on the  zero-dust line) for  each star formation rate  (20 different
  possible ages were used).}  
\vskip 18pt
With  significant flux in  only three  broadband points  (\rb, \ib~and
\zb), it would be  hard to fit each stack to models  with a full range
of every parameter.  However, given what we have learned from studying
these   objects   in  the   color-color   plane,   it  is   relatively
straightforward to fit an age to each of these stacks (assuming a star
formation  history  and  no  dust).   In  order to  fit  the  age,  we
considered 20 different  ages ranging from 1 Myr  - 200 Myr (including
the five ages shown in Figure 2).  By finding the model point (we used
the A\sub{1200} = 0 point  for each model; see discussion for details)
closest to each stack, we have effectively found the average age of an
object in the stack. Because we  didn't want to limit ourselves to one
star formation  history, we have found  two ages for  each object; one
from the closest model with  a constant star formation rate (SFR), and
one from  the closest model with  an exponential SFR.   These ages are
reported in Tables 1 and 2 for constant and exponentially decaying SFR
respectively.  For a  constant SFR, the ages ranged from  1 - 200 Myr,
while for an exponentially decaying SFR, they ranged from 1 - 40 Myr.

In order to find the stellar  mass, we wanted to only use the best-fit
model for each stack.  We took  that to be the model with the best-fit
age found above, and we calculated  a mass for each SFR.  The mass was
found by  a simple weighted  ratio of object  flux to model  flux.  We
derived the weighting by minimizing the ratio of the mass error to the
mass.   The mass  error was  calculated via  a monte  carlo simulation
using  the flux  errors from  each band  for each  stack.   These mass
estimates are reported  in Tables 1 and 2.  The  masses of the objects
from   both  types  of   star  formation   rates  range   from  $\sim$
2$\times$10\super{7} -  2$\times$10\super{9} M\sol.  These  masses are
indicative  of  an average  object  of  the  stack, and  likewise  the
reported error  is the error  in the mean  of the object mass  in each
stack, rather than the error in the measurement.
\vskip 0.25cm
\epsfxsize 3.3in
\epsffile{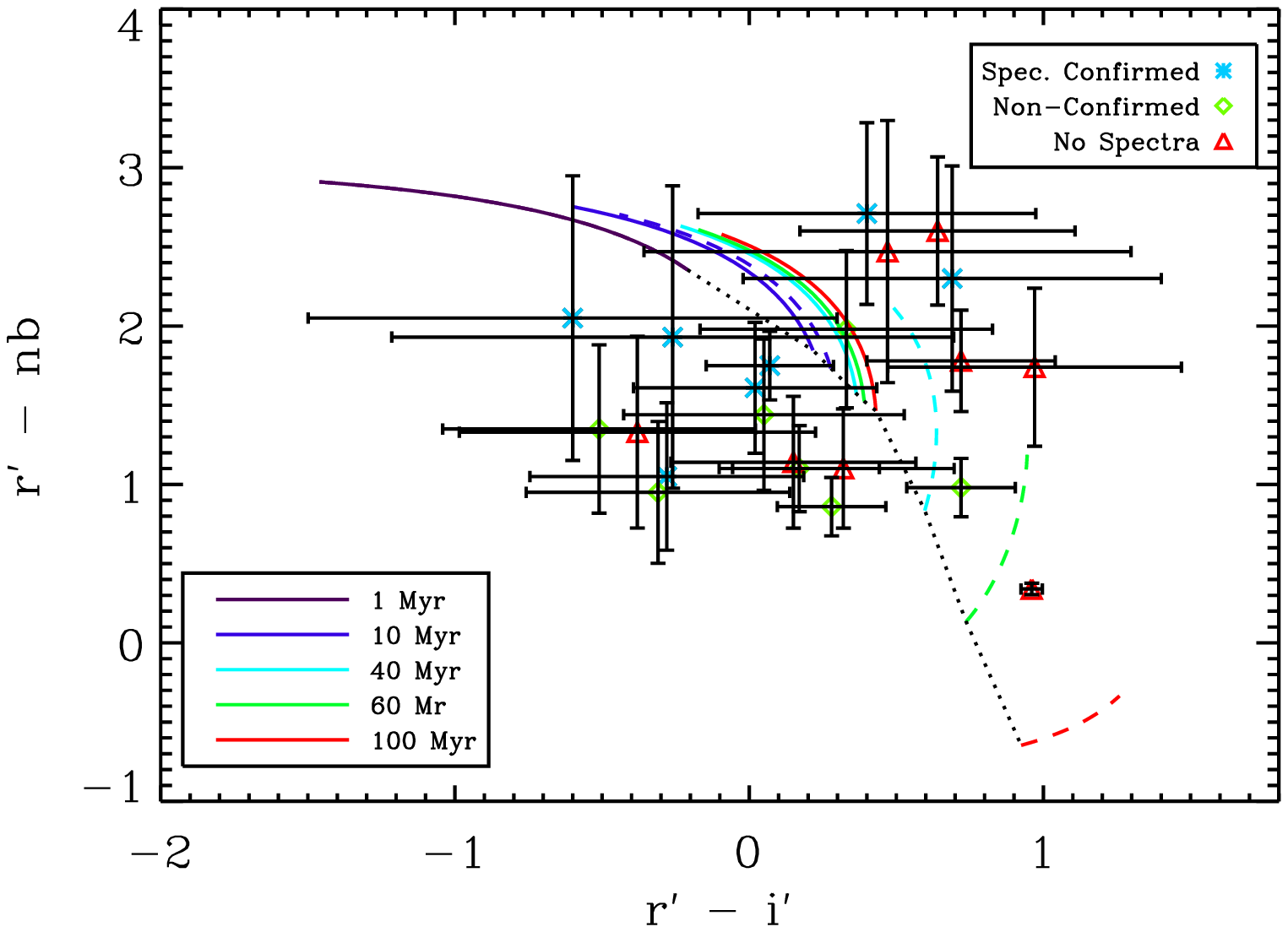}
\figcaption[f3.eps]{Color-color plot  showing the  location of the  21 individual
  detections (along  with the outlier), along  with stellar population
  model tracks.   The individual  galaxies are plotted  with different
  colored symbols  to denote whether they  have been spectroscopically
  confirmed to be  at z $\sim$ 4.5. The axes and  model tracks are the
  same as in  Figure 2.  In general, objects  above the zero-dust line
  would  have their  EW enhanced  due to  dust effects,  while objects
  below this  line would  have intrinsically high  EWs, or  have their
  whole  spectrum  attenuated  by  a  geometrically  homogeneous  dust
  distribution (for the low EW subsample).}
\vskip 18pt
In Figure  3, we  have plotted the  21 individually  detected galaxies
(plus  the one outlier),  along with  their 1$\sigma$~error  bars.  We
have distinguished  between the individual  galaxies in order  to show
those  that were  spectroscopically confirmed.   By their  position in
this  plot, we hope  to find  out whether  they have  an intrinsically
strong \lya~line, or one that is enhanced due to dust effects.

\section{Discussion}

\subsection{Age and Mass Estimates}
We  fitted the  ages of  our  stacked points  using dust-free  models.
Objects in the  stack of the 76 non-detections have  the lowest age by
far  of all of  the stacks  (1 Myr),  and thus  the lowest  mass (0.19
$\times$ 10\super{8} M\sol).  This is not surprising, however, because
these  objects are  selected based  on their  \lya~line  strength.  In
these galaxies, the line is strong  enough to be seen clearly but the
continuum is not, so they are intrinsically faint galaxies, implying a
low stellar mass.  Their colors are  best fit by a low age, which also
points to a low mass; the galaxy  has simply not had that much time to
form stars.

\begin{deluxetable}{ccccc}
\tabletypesize{\small}
\tablecaption{Best-Fit Ages and Masses for a Constant SFR}
\tablewidth{0pt}
\tablehead{
\colhead{Stack} & \colhead{r' mag.} & \colhead{Age} & \colhead{$\chi$\super{2}} & \colhead{Mass (10\super{8}M\sol)}\\
\colhead{$ $} & \colhead{of stack} & \colhead{$(Myr)$} & \colhead{$ $} & \colhead{per object}\\
}
\startdata

7 Low EW&23.27$\pm$0.10&200&18.48&23.54$\pm$1.66\\
7 Mid EW&23.49$\pm$0.12&80&0.93&8.52$\pm$0.74\\
7 High EW&24.08$\pm$0.21&4&0.94&0.95$\pm$0.15\\
21 Det.&22.39$\pm$0.09&40&0.62&4.60$\pm$0.29\\
76 Undet.&23.31$\pm$0.47&1&6.59&0.19$\pm$0.07\\
28 Spec. Conf.&22.96$\pm$0.18&3&0.05&0.62$\pm$0.09\\

\enddata
\tablecomments{Each model has a metallicity of .02 Z\sol.  The low EW bin is from 20--40 \AA; the middle bin is from 45--100 \AA; the highest bin has objects with EW $\geq$ 110 \AA.  The bin with 21 detected objects consists of the 21 galaxies which were detected at the 2$\sigma$ level in at least two of the $r'$, $i'$~or $z'$~bands.  The last bin consists of the 28 objects which were spectroscopically confirmed to have a Ly$\alpha$~line at z $\sim$ 4.5.  The $\chi$\super{2} is defined as the square of the distance in sigma units to the best-fit age point (See Figure 2).  The $r'$~magnitude is for the stack as a whole, while the mass is per object.}

\end{deluxetable}


The three equivalent width bins show  a logical trend in age and mass.
The lowest EW  bin has the highest  age at 200 (40) Myr,  and thus the
highest  mass  at  23.54  (16.15)  $\times$ 10\super{8}  M\sol  for  a
continuous (exponentially decaying) SFR.   This age is consistent with
the fact  that the EWs  in this bin  were not higher than  normal, and
thus  this EW  bin is  best fit  by an  older stellar  population. The
middle EW bin has a younger age and a smaller mass, and the highest EW
bin has the youngest age out of  the three bins at 4 (4) Myr, with the
smallest mass  at 0.95 (0.94) $\times$ 10\super{8}  M\sol.  This would
indicate that the objects with the highest EW are truly the most young
and primitive galaxies in our sample (that were detected). The results
for the  stack of all  21 MMT-detected objects are  intermediate among
the results  grouped by equivalent  width, as expected.  The  ages for
this  stack were  40  (10) Myr,  and  masses were  4.6 (2.0)  $\times$
10\super{8} M\sol.   The stack of  spectroscopically confirmed objects
should lie somewhere between the detected and non-detected stacks, and
indeed this is  the case, with ages  of 3 (2) Myr, and  masses of 0.62
(0.68) $\times$ 10\super{8} M\sol.

The   overall  mass   range  we   found  of   $\sim$   2$\times$10\super{7}  --
2$\times$10\super{9} M\sol  is comparable to other  studies of similar
objects.  Due  to the fact  that our observed magnitudes  are brighter
than the majority of surveys which  we can compare to, we point to the
trend between apparent magnitude and  mass.  That is, that the fainter
an object  is, the less massive it  ought to be (assuming  they are at
similar redshifts).   Ellis et al. (2001) discovered  a very faint  (I $\sim$
30) \lya~emitting  galaxy at z  = 5.576 that  was able to  be detected
because  it was  lensed  by a  foreground  galaxy.  Due  to its  faint
magnitude,  this  object was  determined  to  have  a mass  of  around
10\super{6} M\sol,  implying a  very young age  ($\sim$ 2  Myr).  This
object is consistent with our results, in that it is both much fainter
and much less massive than the  objects we are studying in this paper.
Gawiser et al. (2006)  studied numerous LAEs  which they detected  at z$\sim$
3.1,  with a  median  R magnitude  of  $\sim$ 27.   In  order to  take
advantage of the  full spectral range available to  them, they stacked
their sources before performing SED fitting.  When this was done, they
found  an average  mass per  object of  $\sim$ 5  $\times$ 10\super{8}
M\sol.  All  of our  stacks (except the  stack of  undetected objects)
have an  average individual object  \rb~magnitude of less than  27, so
even  given the differences  in photometric  systems, they  are likely
brighter  (much brighter in  fact, due  to their  increased distance).
While  (depending on  the SFR)  not all  of these  stacks  have masses
higher than 5 $\times$ 10\super{8} M\sol, they are of the same order.

\begin{deluxetable}{cccc}
\tabletypesize{\small}
\tablecaption{Best-Fit Ages and Masses for an Exponentially Decaying SFR}
\tablewidth{0pt}
\tablehead{
\colhead{Stack} & \colhead{Age} & \colhead{$\chi$\super{2}} & \colhead{Mass (10\super{8}M\sol)}\\
\colhead{$ $} & \colhead{$(Myr)$} & \colhead{$ $} & \colhead{per object}\\
}
\startdata

7 Low EW Objects&40&4.05&16.15$\pm$1.14\\
7 Mid EW Objects&20&1.05&4.24$\pm$0.37\\
7 High EW Objects&4&0.97&0.94$\pm$0.14\\
21 Det. Objects&10&3.43&1.97$\pm$0.13\\
76 Undet. Objects&1&6.59&0.19$\pm$0.07\\
28 Spec. Conf. Objects&2&0.09&0.68$\pm$0.10\\

\enddata
\tablecomments{See Table 1 notes.}

\end{deluxetable}


Perhaps the  best comparison  to our results  comes from the  study by
Pirzkal  et al.(2007),  where they  are studying  the properties  of z
$\sim$  5 LAEs  detected in  the GRism  ACS Program  for Extragalactic
Science  (GRAPES)   survey  \citep{grapes}.   Using   models  with  an
exponentially  decaying SFR,  they found  a mass  range of  3 $\times$
10\super{6} -- 3 $\times$  10\super{8} M\sol in objects with an \ib~range
of 25.52 -- 29.35 (the \ib~range of an average object in our stacks is
from  25.12  --  29.90).   While  the majority  of  our  objects  have
\ib~brighter than 26 (all except objects from the undetected stack and
the  spectroscopically confirmed  stack),  all of  the GRAPES  objects
except one have \ib~fainter than 26.  This explains well the fact that
our derived masses  are in general greater than  those derived for the
GRAPES LAEs.   We are studying  intrinsically brighter, and  thus more
massive objects.  While our wavelength  baseline is not very large, we
are able to estimate the mass of stars producing UV and \lya light.

\subsection{Dusty Scenario}
In studying  Figure 3, we  have identified three distinct  regions in
the color-color plane: one region which would house intrinsically blue
objects,  one region  of red  objects  with high  EW, and  one of  red
objects with  low EW.  The region  that lies below  the zero-dust line
and blueward of \rb~- \ib~= 0  would appear to house objects that have
moderate-to-high  EWs  with  blue  colors, indicating  that  they  are
intrinsically blue, containing young stars.  The lower EW/redder \rb~-
\ib~color sub-area of this region  would contain objects may have some
dust, but it  is affecting both the continuum  and \lya~line, lowering
the  EW and reddening  the color.

The second region  consists of the area above  the zero-dust line with
an \rb~- $nb$ color $\gtrsim$  2.  This region would hold objects that
lie  in the  dusty regime  of the  models, yet  they would  still have
``red'' \rb~-  $nb$ colors,  indicating a large  EW.  Objects  such as
these could be  explained by dust quenching of  the continuum (causing
the red \rb~- \ib~color), while the \lya~line would not be affected by
the dust,  making it appear to  be strong (causing the  red \rb~- $nb$
color).   The last region  consists of  the area  above the  zero dust
line, but with a smaller \rb~-  $nb$ color (\rb~- \ib~ $\sim$ 1, \rb~-
$nb$  = 1 --  1.5).  Objects  in this  area would  have a  redder \rb~-
\ib~color, with  a smaller \rb~-  $nb$ color (smaller  EW), indicating
that they are perhaps older galaxies with some dust effects present.

In  the above  paragraph, we  have outlined  different regions  in our
color-color  plane  which  would  house  objects  from  the  different
scenarios we are examining.  We plan to use this to determine the
likelihood of  dust quenching  of the continuum  resulting in  a large
\lya~EW.  However, the current error bars on individual
objects are large compared to  their distance from the zero-dust line.
In order to  quantify this, we calculated the  distance of each object
from the  zero-dust line in units  of $\sigma$.  The  mean distance of
the 21 objects we are studying  is 1.10 $\pm$ 0.48 $\sigma$.  While it
is possible that  the dust enhancement of the  \lya~line is present in
our  sample, due  to  the large  error  bars we  cannot  rule out  the
possibility that all  of these objects lie on the  zero-dust line at a
level of $\sim$ 1 $\sigma$.

Performing the  same exercise with  the six stacked points  with their
smaller error bars, the mean  distance from the zero-dust line is 1.17
$\pm$ 0.81 $\sigma$.   However, the smaller error bars  do allow us to
make more of  a characterization of a few of the  stacks.  We can most
likely rule out  clumpy dust enhancement of the  \lya~line for the low
and  middle EW  stacks, mainly  because they  lie below  the zero-dust
line, and far from the region  in our color-color plane where we would
expect  to find  the enhanced  line.   Also, in  the line  enhancement
scenario,  one would  expect to  find extremely  high EW \lya~lines, and
these two  stacks do not  contain the highest  33\% EW objects  in our
sample.   There still may  be dust  present in  these objects,  but it
would be in  a uniform distribution, quenching both  the continuum and
the \lya~line.

\section{Conclusion}

We have  used MMT/Megacam broadband  photometry in order to  study the
continuum  properties of  \lya~emitting galaxies  at z$\sim$  4.5.  By
dividing our objects into six different catergories and stacking them,
we were able to derive age and stellar mass estimates for an average galaxy in
each stack (ignoring  dust effects).  In all cases,  the best fit ages
were young,  with the oldest age  coming from objects  with the lowest
equivalent width.  Even  those objects had an age of  only 200 Myr (40
Myr) for  a continuous  (exponentially decaying) star  formation rate.
As would be expected, the bin with the highest EW had the youngest age
out of the continuum detected objects  at 4 Myr (from both SFRs).  The
youngest age from both SFRs came from objects whose continuum flux was
undetected,  mostly  because these  were  detected  in the  narrowband
image, and not in the continuum images.

The  derived stellar masses ranged  from  $\sim$  2$\times$10\super{7}  M\sol for  the
undetected objects to $\sim$ 2$\times$10\super{9} for the lowest EW
objects.  Our derived masses  are consistent with other mass estimates
of  similar objects.   In the  majority of  cases, the  masses  of our
objects were greater than those from other studies (Ellis et al. 2001;
Gawiser et al.  2006; Pirzkal et al.  2007).   However, the magnitudes
of our individual  objects were brighter than those  of the comparison
studies, implying  that our objects are larger,  more massive galaxies
than those from  the other studies.  In conclusion,  while we have not
yet  been  able  to  definitively  determine the  likelihood  of  dust
enhancement of the \lya~line, we have  been able to shed some light on
the physical properties of these high-z objects.  The \lya~galaxies in
this survey,  especially those with  the largest \lya~EW, are  some of
the youngest and least-massive objects  in the early universe known to
date.

\begin{acknowledgements}
  This work  was supported by the Arizona  State University/NASA Space
  Grant, the ASU Department of Physics and the ASU School of Earth and
  Space  Exploration.   This  work  made  use of  images  and/or  data
  products provided  by the NOAO  Deep Wide-Field Survey  (Jannuzi and
  Dey  1999), which  is supported  by the  National  Optical Astronomy
  Observatory  (NOAO).   NOAO  is  operated  by AURA,  Inc.,  under  a
  cooperative agreement with the National Science Foundation.
\end{acknowledgements}


\begin{thebibliography}{}
\bibitem[Ajiki(2004)]{ajiki04} Ajiki, M. et al. 2004, \pasj\ 56, 597
\bibitem[Bertin  \&   Arnouts(1996)]{ba96}  Bertin,  E.   \&  Arnouts, S. 1996, A\&AS, 117, 393
\bibitem[Bruzual \& Charlot(2003)]{bc03} Bruzual, G. \& Charlot, S. 2003, MNRAS, 344, 1000
\bibitem[Calzetti et al.(1994)]{cal} Calzetti, D., Kinney, A. L. \& Storchi-Bergmann, T. 1994, ApJ, 429, 582
\bibitem[Charlot \& Fall(1993)]{cf93} Charlot, S. \& Fall, S. M. 1993, \apj, 415, 580
\bibitem[Cowie \& Hu(1998)]{cow98} Cowie, L. L., \& Hu, E. M. 1998, \aj, 115, 1319
\bibitem[Dawson et al.(2004)]{daw04} Dawson, S., Rhoads, J. E., Malhotra, S., Stern, D., Dey, A., Spinrad, H., Jannuzi, B. T., Wang, J. X. \& Landes, E. 2004, \apj, 617, 707

\bibitem[Ellis et al.(2001)]{ell01} Ellis, R., Santos, M. R., Kneib, J.-P. \& Kuijken, K. 2001, \apj, 560, L119 
\bibitem[Fujita et al.(2003)]{fujita03} Fujita, S. S. et al. 2003, \apj, 586, L115
\bibitem[Fukugita et al.(1996)]{fuk} Fukugita, M., Ichikawa, T., Gunn, J. E., Doi, M., Shimasaku, K. \& Schneider, D. P. 1996, AJ, 111, 1748
\bibitem[Fynbo et al.(2001)]{fynbo} Fynbo, J. U., M$\ddot{o}$ller, P. \& Thomsen, B. 2001, A\&A, 374, 443
\bibitem[Gawiser et al.(2006)]{gaw06} Gawiser, E. et al. 2006, /apj, 642, L13
\bibitem[Hansen \& Oh(2006)]{hanoh} Hansen, M. \& Oh, S. P. 2006, MNRAS, 367, 979
\bibitem[Hu et al.(1998)]{hu98} Hu, E. M., Cowie, L. L. \& McMahon, R. G. 1998, \apj, 502, L99
\bibitem[Hu et al.(2002)]{hu02} Hu, E. M., Cowie, L. L., McMahon, R. G., Capak, P., Iwamuro, F., Kneib, J.-P., Maihara, T. \& Motohara, K. 2002, ApJ, 568, L75
\bibitem[Hu et al.(2004)]{hu04} Hu, E. M., Cowie, L. L., Capak, P., McMahon, R. G., Hayashino, T. \& Komiyama, Y. 2004, AJ, 127, 563
\bibitem[Jannuzi \& Dey(1999)]{jdey99} Jannuzi, B. T., Dey, A. 1999, in ASP Conf. Ser. 191, Photometric Redshifts and High Redshift Galaxies, ed. R. J. Weymann, L. J. Storrie-Lombardi, M. Sawicki, \& R. J. Brunner (San Francisco: ASP), 111
\bibitem[Kodaira et al.(2003)]{kod03} Kodaira, K. et al. 2003, PASJ, 55, L17
\bibitem[Kudritzki et al.(2000)]{kud00} Kudritzki, R.-P. et al. 2000, ApJ, 536, 19
\bibitem[Landolt(1992)]{lan} Landolt, A. U. 1992, AJ, 104, 340
\bibitem[Madau(1995)]{mad} Madau, P. 1995, ApJ, 441, 18
\bibitem[Malhotra \& Rhoads(2002)]{mr02} Malhotra, S. \& Rhoads, J. E. 2002, ApJ, 565, L71
\bibitem[Malhotra et al.(2003)]{mal03} Malhotra, S., Wang, J.X., Rhoads, J. E., Heckman, T. M. \& Norman, C. A. 2003, \apj, 585, L25
\bibitem[Malhotra \& Rhoads(2004)]{mr04} Malhotra, S. \& Rhoads, J. E. 2004, ApJ, 617, L5
\bibitem[McLeod et al.(2006)]{obs}McLeod, B. M., Caldwell, N., Williams, G. \& Conroy, M. 2006, Megacam Observers Manual, http://grb.mmto.arizona.edu/$\sim$ggwilli/mmt/docs/megacam/\\megacam$\_$manual.html
\bibitem[McLeod et al.(1998)]{mega} McLeod,  B. A.,  Gauron, T. M., Geary,  J. C., Ordway, M.  P. \& Roll, J. B.  1998, SPIE, 3355, 477
\bibitem[Neufeld(1991)]{neufeld} Neufeld, D. A. 1991, \apj, 370, L85
\bibitem[Oke \& Gunn(1983)]{oke} Oke, J. B. \& Gunn, J. E. 1983, \apj, 266, 713
\bibitem[Ouchi et al.(2001)]{ouchi01} Ouchi, M. et al. 2001, ApJ, 558, 83
\bibitem[Ouchi et al.(2003)]{ouchi03} Ouchi, M. et al. 2003, ApJ, 582, 60
\bibitem[Ouchi et al.(2004)]{ouchi04} Ouchi, M. et al. 2004, ApJ, 611, 685
\bibitem[Papovich et al.(2001)]{pap} Papovich, C., Dickinson, M. \& Ferguson, H. C. 2001, ApJ, 559, 620
\bibitem[Partridge \& Peebles(1967)]{pp67} Partridge, R. B. \& Peebles, P. J. E. 1967, \apj, 147, 868
\bibitem[Pentericci et al.(2000)]{pen00} Pentericci, L. et al. 2000, \aap, 361, 25
\bibitem[Pirzkal et al.(2004)]{grapes} Pirzkal, N. et al. 2004, ApJS, 154, 501
\bibitem[Pirzkal et al.(2007)]{pirz06} Pirzkal, N. et al. 2007, in preparation
\bibitem[Rhoads(2000)]{jrpasp} Rhoads, J. E. 2000, PASP, 112, 771, 703
\bibitem[Rhoads et al.(2000)]{rh00} Rhoads, J. E., Malhotra, S., Dey, A., Stern, D., Spinrad, H. \& Jannuzi, B. T. 2000, ApJ, 545, L85
\bibitem[Rhoads \& Malhotra(2001)]{rho01} Rhoads, J. E. \& Malhotra, S. 2001, \apj\ 563, L5 
\bibitem[Rhoads et al.(2003)]{rho03} Rhoads, J. E., Dey, A., Malhotra, S., Stern, D., Spinrad, H., Jannuzi, B. T., Dawson, S., Brown, M. J. I. \& Landes, E. 2003, AJ, 125, 1006
\bibitem[Rhoads et al.(2004)]{rho04} Rhoads, J. E., Xu, C., Dawson, S., Dey, A., Malhotra, S., Wang, J. X., Jannuzi, B. T., Spinrad, H., \& Stern, D. 2004, \apj, 611, 59
\bibitem[Shimasaku et al.(2003)]{shima03} Shimasaku, K. et al. 2003, \apj, 586, L111
\bibitem[Shimasaku et al.(2006)]{shima06} Shimasaku, K. et al. 2006, \pasj,58, 313
\bibitem[Steidel et al.(1996)]{steidel} Steidel, C. C., Giavalisco, M., Pettini, M., Dickinson, M. \& Adelberger, K. L. 1996, \apj, 462, L17
\bibitem[Stiavelli et al.(2001)]{stia01} Stiavelli, M.,  Scarlata, C., Panagia, N., Treu, T., Bertin, G. \& Bertola, F. 2001, \apj, 561, L37
\bibitem[Taniguchi et al.(2005)]{tani05} Taniguchi, Y. et al. 2005, \pasj, 57, 165
\bibitem[Tody(1993)]{tody93} Tody, D. 1993, ASPC, 52, 173
\bibitem[Tody(1986)]{tody86} Tody, D. 1986, SPIX, 627, 733
\bibitem[Valdes(1998)]{val98} Valdes, F. G. 1998, ASPC, 145, 53
\bibitem[Valdes \& Tody(1998)]{valtod} Valdes, F. G. \& Tody, D. 1998, SPIE, 3355, 497
\bibitem[Venemans et al.(2002)]{vene02} Venemans, B. P. et al. 2002, \apj, 569, L11
\bibitem[Venemans et al.(2004)]{vene04} Venemans, B. P. et al. 2004, \aap, 424, L17
\bibitem[Wang et al.(2004)]{wang04} Wang, J. X., Rhoads, J. E., Malhotra, S., Dawson, S., Stern, D., Dey, A., Heckman, T. M., Norman, C. A. \& Spinrad, H. 2004, \apj, 608, L21
\bibitem[Wang et al.(2007)]{wang07} Wang, J. X. et al. 2007, in preparation

\end{thebibliography}
\end{document}